\documentclass[twocolumn,showpacs,preprintnumbers,superscriptaddress,showkeys,pra]{revtex4-1}
\usepackage{dcolumn}
\usepackage{hyperref}

\usepackage{amsmath,bbm}
\usepackage{amssymb}
\usepackage{graphicx}
\usepackage{mathrsfs}
\usepackage{color}

\newcommand{\ctr}{{x_c}}
\newcommand{\ovp}{\chi}
\newcommand{\pwr}{{\mathscr{P}}}
\newcommand{\snr}{{\cal{S}}}
\newcommand{\xrf}{{x_r}}
\newcommand{\cur}{{\mathscr{I}}}
\newcommand{\noi}{{\mathscr{N}}}
\newcommand{\Der}{\textrm{d}}

\DeclareGraphicsExtensions{.png,.pdf,.eps,.jpg}
\begin{document}

\title{Sub-Rayleigh resolution of two incoherent sources by array homodyning}

\author{Chandan Datta}
\email{c.datta@cent.uw.edu.pl}
\affiliation{Centre for Quantum Optical Technologies, Centre of New Technologies, University of Warsaw, Banacha 2c, 02-097 Warszawa, Poland}

\author{Marcin Jarzyna}
\email{m.jarzyna@cent.uw.edu.pl}
\affiliation{Centre for Quantum Optical Technologies, Centre of New Technologies, University of Warsaw, Banacha 2c, 02-097 Warszawa, Poland}

\author{Yink Loong Len}
\email{y.len@cent.uw.edu.pl}
\affiliation{Centre for Quantum Optical Technologies, Centre of New Technologies, University of Warsaw, Banacha 2c, 02-097 Warszawa, Poland}

\author{Karol {\L}ukanowski}
\email{k.lukanowski@cent.uw.edu.pl}
\affiliation{Centre for Quantum Optical Technologies, Centre of New Technologies, University of Warsaw, Banacha 2c, 02-097 Warszawa, Poland}
\affiliation{Faculty of Physics, University of Warsaw, Pasteura 5, 02-093 Warszawa, Poland}

\author{Jan Ko{\l}ody\'{n}ski}
\email{jan.kolodynski@cent.uw.edu.pl}
\affiliation{Centre for Quantum Optical Technologies, Centre of New Technologies, University of Warsaw, Banacha 2c, 02-097 Warszawa, Poland}

\author{Konrad Banaszek}
\email{k.banaszek@uw.edu.pl}
\affiliation{Centre for Quantum Optical Technologies, Centre of New Technologies, University of Warsaw, Banacha 2c, 02-097 Warszawa, Poland}
\affiliation{Faculty of Physics, University of Warsaw, Pasteura 5, 02-093 Warszawa, Poland}

\begin{abstract}

Conventional incoherent imaging based on
measuring the spatial intensity distribution in the image plane faces the resolution hurdle described by the Rayleigh diffraction criterion.
Here, we demonstrate theoretically using the concept of the Fisher information
that quadrature statistics measured by means of array homodyne detection enables estimation of the distance between two incoherent point sources well below the Rayleigh limit for sufficiently high signal-to-noise ratio. This capability is attributed to the availability of spatial coherence information between individual detector pixels acquired using the coherent detection technique. A simple analytical approximation for the precision attainable in the sub-Rayleigh region is presented. Furthermore, an estimation algorithm is proposed and applied to Monte Carlo simulated data.
\end{abstract}

\maketitle

\section{Introduction}

While the optical
band enables insightful observations of physical, chemical, and biological systems, resolving their spatial characteristics is often hindered
by diffractive limitations of imaging instruments described by the fundamental Rayleigh criterion \cite{Rayleigh1879}. The Rayleigh limit follows from the direct detection of the spatial intensity distribution of the optical field in the image plane. The purpose of this paper is to identify theoretically the capability of array homodyne detection \cite{FinkAO1976,LeClercOL2000,BeckPRL2000} to resolve sub-Rayleigh features in optical imaging. The crucial benefit of array homodyning is the availability of the spatial coherence information for the optical field detected in the image plane. Such coherence is introduced by the transfer function of the imaging system between the source and the image planes even if contributions from individual points
constituting the source
are mutually incoherent. This observation underlies currently explored approaches to overcome the Rayleigh limit by detecting the optical field in a carefully selected basis of spatial modes in the image plane \cite{TsangPRX2016,LupoPRL2016,PaurOPT2016,YangOPT2016,YangPRA2017,LarsonOPT2018,HradilOPT2019}. The scenario considered here will assume that individual pixels of the homodyne detector have dimension much smaller than the spatial variation of the transfer function, but contribute noise that is independent of their size. This noise model includes the important case of homodyne detection operated at the ultimate shot-noise limit. The sub-Rayleigh sensitivity of array homodyning will be demonstrated for the canonical example of a binary source, where light is emitted by two equally bright and mutually incoherent points.

The remainder of this paper is organized as follows. In Sec.~\ref{Sec:ImagingSystem} we describe the model of the imaging system with array homodyne detection. Sec.~\ref{Sec:Fisher} reviews the concept of the Fisher information matrix in the context of the measurement scheme under consideration. Precision of estimating the separation and the centroid of a binary source from homodyne measurements is analyzed in Sec.~\ref{Sec:Precision}. A practical estimation procedure is described and tested using Monte Carlo simulated homodyne data in Sec.~\ref{Sec:EstProcedure}. Finally, Sec.~\ref{Sec:Conclusions} concludes the paper.

\section{Imaging system}
\label{Sec:ImagingSystem}

For simplicity, we shall consider a one-dimensional model of the imaging system characterized by a normalized transfer function $u(x)$, where $x$ parametrises the transverse spatial coordinate in the image plane.
Let the source be in general composed of a finite number of points
emitting quasi-monochromatic light described by thermal statistics. The electromagnetic field in the image plane is then represented by the complex signal
\begin{equation}
\mathscr{E}(x) = \sum_{l} \alpha_l u_l (x),
\end{equation}
where the $l$th source contributes the displaced transfer function $u_l (x)= u(x- x_l)$ multiplied by an amplitude $\alpha_l$ characterized by a complex normal distribution $\alpha_l \sim {\cal CN}(0, \pwr w_l)$ with a zero mean. For convenience, the variance of $\alpha_l$ is expressed as a product of
the total optical power
\begin{equation}
\pwr = \int_{-\infty}^{\infty} \mathbb{E}[|\mathscr{E}(x)|^2] \, \Der x
\end{equation}
reaching the image plane and the relative weights $w_l$ of contributions from individual sources that add up to one, $\sum_l w_l =1$.

\begin{figure}
\centering
\includegraphics[width=0.95\columnwidth]{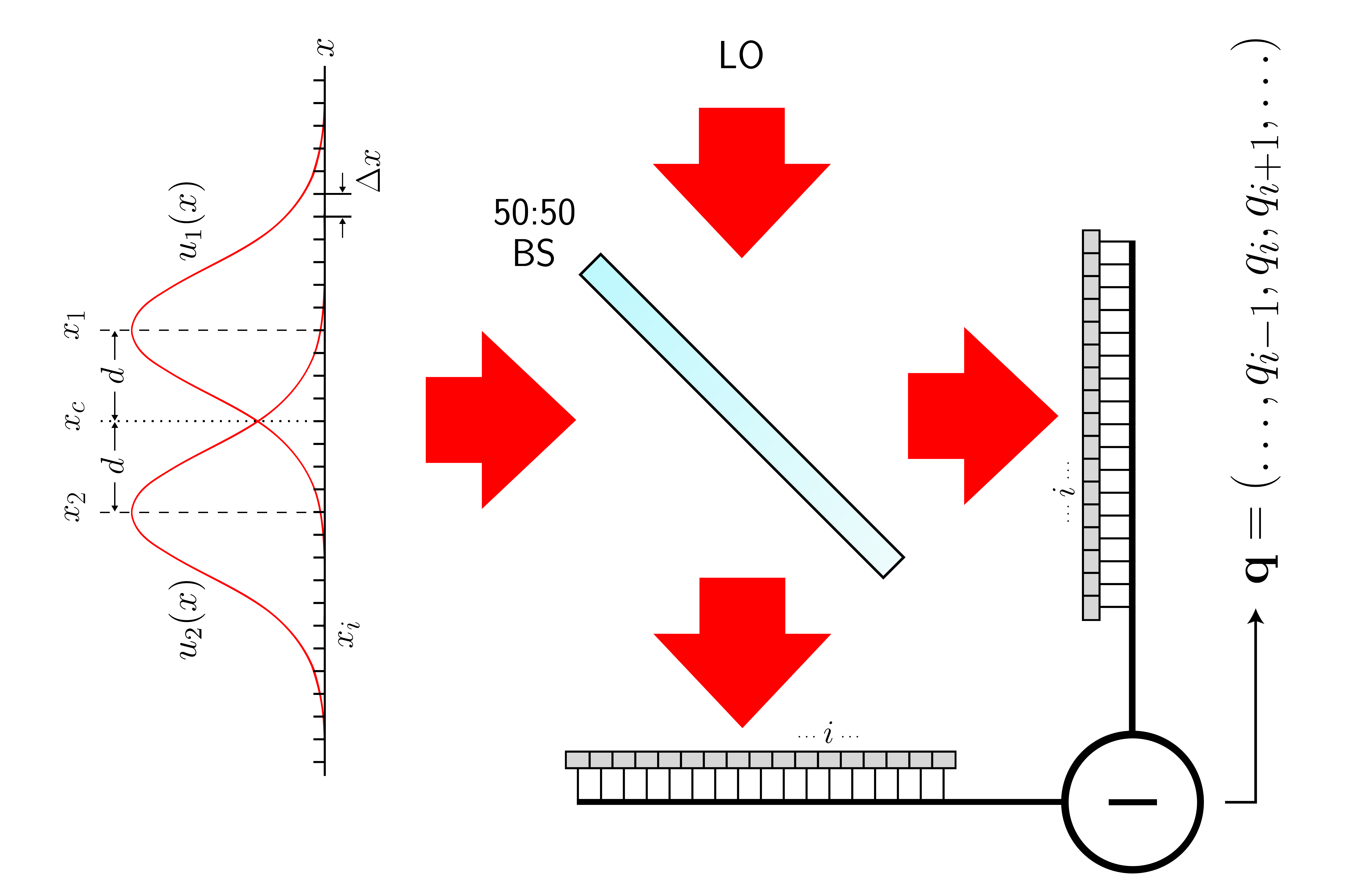}
\caption{(Color online) Balanced homodyne detection in the image plane. For concreteness, a field produced by a binary source with two components $u_1(x)=u(x-x_1)$ and $u_2(x)=u(x-x_2)$ separated by $2d=x_1-x_2$ and centered at $\ctr=(x_1+x_2)/2$ is schematically shown. The signal field is superposed with a strong, flat-wavefront local oscillator LO on a 50:50 beam splitter BS with outputs monitored by array detectors. Subtracting photocurrents in between pairs of matching individual pixels yields after rescaling the quadrature vector $\mathbf{q}=(\ldots,q_{i-1},q_i,q_{i+1}, \ldots)$.}
\label{Fig:Schematic}
\end{figure}

As shown in Fig.~\ref{Fig:Schematic}, the field in the image plane is detected by means of balanced array homodyne detection with uniform both the efficiency and the noise figure across all the pixels. The $i$th detector pixel pair is centered at $x_i$ and their size $\Delta x$ is assumed to be much smaller than the spatial variation of the transfer function. The differential photocurrent measured between the $i$th pair of pixels
is a sum of contributions from the optical signal $\cur_i^{\text{s}}$ and the detector noise $\cur_i^{\text{n}}$, the latter assumed to be Gaussian with a zero mean and uncorrelated between pixels. If the local oscillator phase is uniform  across the array and set to zero, the signal contribution reads $\cur_i^{\text{s}} = \sqrt{\Delta x/2} [\mathscr{E}(x_i)+\mathscr{E}^\ast(x_i)]$. The detection noise variance $\mathrm{Var}[\cur_i^{\text{n}}] = \noi$
is taken as independent of the pixel size.
In such a case it is convenient to rescale the quadrature measured at the $i$th pixel pair as $q_i = (\cur_i^{\text{s}} + \cur_i^{\text{n}})/\sqrt{\noi}$. The quadrature vector $\mathbf{q} = (\ldots, q_{i-1}, q_i, q_{i+1}, \ldots)^T$ is then characterized by a multivariate normal distribution with a zero mean, ${\mathbbm E}({\mathbf q})=0$, and the covariance matrix
\begin{equation}
\label{Eq:Covariance}
\mathbf{C} = {\mathbbm E} [ {\mathbf q} {\mathbf q}^T ]  = \snr\, \boldsymbol{\Gamma} +  \mathbf{I} ,
\end{equation}
where $\snr = \pwr/\noi$ is the signal-to-noise ratio (SNR) and $\mathbf{I}$ stands for the identity matrix. For an infinitesimal pixel size, the elements of the matrix $\mathbf{\Gamma}$ can be written as
\begin{equation}
\Gamma_{ii'} =  \text{Re}[\Gamma(x_i, x_{i'})]\Delta x,
\end{equation}
where $\Gamma(x, x')$ is the coherence function normalized by the total signal power in the image plane,
\begin{equation}
\label{Eq:CoherenceFunction}
\Gamma(x, x')= \frac{1}{\pwr} \mathbbm{E} [\mathscr{E}^\ast(x)\mathscr{E}(x') ] =
\sum_{l} w_l u_l^\ast(x)u_l (x').
\end{equation}
The second expression given above follows directly from the absence of coherence between individual point sources.
In the case of shot-noise-limited homodyning, the SNR reads twice the total average photon number reaching the image plane from the source.

In the following, we will consider a binary source comprising two equally bright points producing contributions $u_1(x)=u(x-x_1)$ and $u_2(x) = u(x-x_2)$ in the image plane located respectively at $x_1 = \ctr +d$ and $x_2 = \ctr -d$, as shown in Fig.~\ref{Fig:Schematic}. Here $d$ specifies the half-separation between the points and $\ctr$ is the centroid of this binary source. Two models of the
normalized real transfer function will be used in numerical examples:
\begin{equation}
\label{Eq:trans_funs}
u(x)=
\begin{cases}
(2/\pi\sigma^{2})^{1/4}\exp(-x^{2}/\sigma^{2}), & \text{[soft aperture]}\\
3^{1/4}/ \sqrt{\pi\sigma} \textrm{sinc}(\sqrt{3}x/\sigma) . & \text{[hard aperture]}
\end{cases}
\end{equation}
In both cases,
the first derivative $u'(x)$ of the transfer function is used to characterize its spatial spread as
\begin{equation}
\label{Eq:sigmadef}
\sigma =  \left( \int_{-\infty}^{\infty}
[u'(x)]^2
\Der x \right)^{-1/2},
\end{equation}
which in turn defines for regular transfer functions the Rayleigh limit
below which the resolution of conventional direct imaging is lost \cite{YinkIJQI2020}.

\section{Fisher information matrix}
\label{Sec:Fisher}

The precision of estimating the source parameters will be quantified using the concept of the Fisher information (FI) matrix. Consider a general scenario where a sample of $N$ vectors composed of real random variables $\mathbf{q}= (\ldots, q_{i-1}, q_i, q_{i+1}, \ldots)^T$ is used to determine values of parameters $\theta_j$ with the help of estimators $\tilde{\theta}_j$ that are unbiased, i.e.\ $\mathbbm{E}[\tilde{\theta}_j] = \theta_j$. The estimation precision can be characterized by the covariance matrix ${\bm{\mathcal C}}$  with elements ${\mathcal C}_{jj'} = \textrm{Cov}[\tilde{\theta}_j, \tilde{\theta}_{j'}]$. This covariance matrix satisfies the Cram\'{e}r-Rao bound
\begin{equation}
{\boldsymbol{\mathcal C}} \ge (N{\boldsymbol{\mathcal F}})^{-1},
\end{equation}
where ${\boldsymbol{\mathcal F}}$ is the Fisher information matrix. Note that the variance of individual estimators
is lower bounded by $\mathrm{Var}[\tilde{\theta}_j] \ge ({\boldsymbol{\mathcal F}}^{-1})_{jj} \ge
({\mathcal F}_{jj})^{-1}$. The second inequality is not necessarily tight when the Fisher information matrix is not diagonal.

As discussed in Sec.~\ref{Sec:ImagingSystem}, for homodyne detection of thermal sources the variables $\mathbf{q}$ follow a normal multivariate distribution with a covariance matrix $\mathbf{C}$ given by Eq.~(\ref{Eq:Covariance}) and a zero mean, ${\mathbbm E}[\mathbf{q}]=0$. In this case, individual elements of the Fisher information matrix ${\boldsymbol{\mathcal F}}$ are given by \cite{slepian1954estimation, bangs1972array}:
\begin{equation}
\label{Eq:FIMNormal}
{\cal F}_{jj'} =
\frac{1}{2}\mbox{Tr}\left( {\mathbf{C}}^{-1} \frac{\partial {\mathbf{C}}}{\partial\theta_j}
{\mathbf{C}}^{-1} \frac{\partial {\mathbf{C}}}{\partial\theta_{j'}}
\right).
\end{equation}
When the number of points emitting radiation in the source plane is finite, the matrix $\boldsymbol\Gamma$
appearing in Eq.~(\ref{Eq:Covariance}) has a finite decomposition into non-zero eigenvalues $\gamma_\mu$ and corresponding eigenvectors ${\mathbf{e}}_\mu$ of the form
\begin{equation}
{\boldsymbol\Gamma} = \sum_{\mu} \gamma_\mu {\mathbf{e}}_\mu {\mathbf{e}}_\mu^T.
\end{equation}
This allows one to write the
covariance matrix $\mathbf{C}$ as a finite sum:
\begin{equation}
\label{Eq:Covform}
\mathbf{C} = \sum_\mu V_\mu \mathbf{e}_\mu \mathbf{e}_\mu^T + \mathbf{P}_\perp.
\end{equation}
The orthonormal eigenvectors $\mathbf{e}_\mu$ determine principal components distinguished by the fact that their variances $V_\mu=\mathrm{Var}[\mathbf{e}_\mu^T \mathbf{q}]=\snr \gamma_\mu+1 $ differ from one, $V_\mu \neq 1$. The matrix $\mathbf{P}_\perp = \mathbf{I} - \sum_\mu \mathbf{e}_\mu\mathbf{e}_\mu^T$ is the projection onto the subspace orthogonal to all the eigenvectors $\mathbf{e}_\mu$.

For the covariance matrix of the form given by Eq.~(\ref{Eq:Covform}), the Fisher information matrix evaluated according to Eq.~(\ref{Eq:FIMNormal}) can be written as a sum of three contributions
\begin{equation}
{\boldsymbol{\mathcal F}} = {\boldsymbol{\mathcal F}}^{(1)} + {\boldsymbol{\mathcal F}}^{(2)}
+ {\boldsymbol{\mathcal F}}^{(3)}
\end{equation}
that result respectively from the change with the parameters $\theta_j$ of the principal component variances $V_\mu$, of the eigenvectors $\mathbf{e}_\mu$ in the orthogonal subspace $\mathbf{P}_\perp$, and of the eigenvectors $\mathbf{e}_\mu$ in the subspace spanned by their set. The explicit expressions for the three contributions are derived in Appendix~\ref{App:A} as Eqs.~(\ref{Eq:F(1)jj'})--(\ref{Eq:F(3)jj'}). These expressions involve scalar products between the eigenvectors $\mathbf{e}_\mu$ and/or their derivatives
$\partial\mathbf{e}_\mu/\partial{\theta_j}$ with respect to estimated parameters $\theta_j$. For the small pixel size considered here, the eigenvectors $\mathbf{e}_\mu$ will be effectively given by values of real continuous functions $e_\mu(x)$ on a grid $\ldots, x_{i-1}, x_{i}, x_{i+1}, \ldots$ with spacing $\Delta x$ according to
\begin{equation}
\label{Eq:DiscreteVectors}
\mathbf{e}_\mu = \sqrt{\Delta x} \bigl( \ldots,e_\mu(x_{i-1}) ,e_\mu(x_{i}), e_\mu(x_{i+1}), \ldots \bigr)^T.
\end{equation}
Then the relevant scalar products appearing in Eqs.~(\ref{Eq:F(1)jj'})--(\ref{Eq:F(3)jj'}) can be expressed as integrals, e.g.\ $\mathbf{e}_\mu^T \partial \mathbf{e}_{\mu'}/\partial {\theta_j} = \int_{-\infty}^{\infty}   e_\mu(x)  [\partial e_{\mu'}(x)/\partial{\theta_j}] \Der x$, provided that the functions $e_\mu(x)$ vary slowly over the scale defined by $\Delta x$.

\section{Estimation precision}
\label{Sec:Precision}

The general formalism presented in the preceding section can be now applied to
characterize the precision of estimating the half-separation $d$ and the centroid $x_c$ of a binary source from measured quadrature vectors $\mathbf{q}$. The distinguished eigenvectors appearing in Eq.~(\ref{Eq:Covform})
can be found in the continuous limit from the Karhunen--Lo\`{e}ve decomposition \cite{KLDecomposition} of the coherence function $\Gamma(x,x')$ defined in Eq.~(\ref{Eq:CoherenceFunction}), which for the binary source takes the form
\begin{equation}
\Gamma(x, x') =  \gamma_- e_-^\ast(x)e_- (x') +
\gamma_+ e_+^\ast(x)e_+ (x').
\end{equation}
In the case of two equally bright points and a real-valued transfer function the
two eigenvalues read
\begin{equation}
\label{Eq:gammapm}
\gamma_\pm = \frac{1}{2}(1\pm \ovp),
\end{equation}
where
\begin{equation}
\label{Eq:OverlapDef}
\ovp= \int_{-\infty}^{\infty}  u_1(x) u_2(x) \, \Der x
\end{equation}
is the overlap between the two displaced transfer functions $u_1(x)=u(x-\ctr-d)$ and $u_2(x)=u(x-\ctr+d)$ corresponding to individual points in the binary source. The respective normalized eigenmodes are real and given explicitly by
\begin{equation}
\label{Eq:epm(x)}
e_\pm (x) = \frac{u_1(x) \pm u_2(x)}{\sqrt{2(1\pm\ovp)}}.
\end{equation}
The two distinguished eigenvectors $\mathbf{e}_\pm$ of the covariance matrix $\mathbf{C}$
can be written in a normalized form using the values of the eigenmodes $e_\pm(x)$ on the pixel grid
according to Eq.~(\ref{Eq:DiscreteVectors}).
These vectors define two principal components $q_\pm = \mathbf{e}_\pm^T \mathbf{q}$ of the multivariate quadrature distribution that are characterized by a zero mean and respective variances:
\begin{equation}
\label{Eq:Vpm}
V_\pm = \textrm{Var}[q_\pm] = \snr \gamma_{\pm} + 1 = \snr(1\pm \ovp)/2+1,
\end{equation}
where in the third step we have used Eq.~(\ref{Eq:gammapm}).
The variances of all other
components of $\mathbf{q}$ that are orthogonal to $\mathbf{e}_\pm$ are equal to one.

In the following, for notational simplicity it will be convenient to use $d$ and $\ctr$ as the indices of the two-dimensional Fisher information matrix.
Moreover, it can be verified by a direct calculation that in the case considered here the off-diagonal element of the Fisher information matrix vanishes, ${\cal F}_{\ctr d}=0$. Hence the estimation of the half-separation and the centroid can be treated as statistically independent and one can use a single index $j=d, \ctr$ to label diagonal elements of the Fisher information matrix, ${\cal F}_{j} \equiv {\cal F}_{jj}$.

\subsection{Half-separation estimation}
\label{Sec:SeparationPrecision}

\begin{figure}
\centering
\includegraphics[width=0.9\columnwidth]{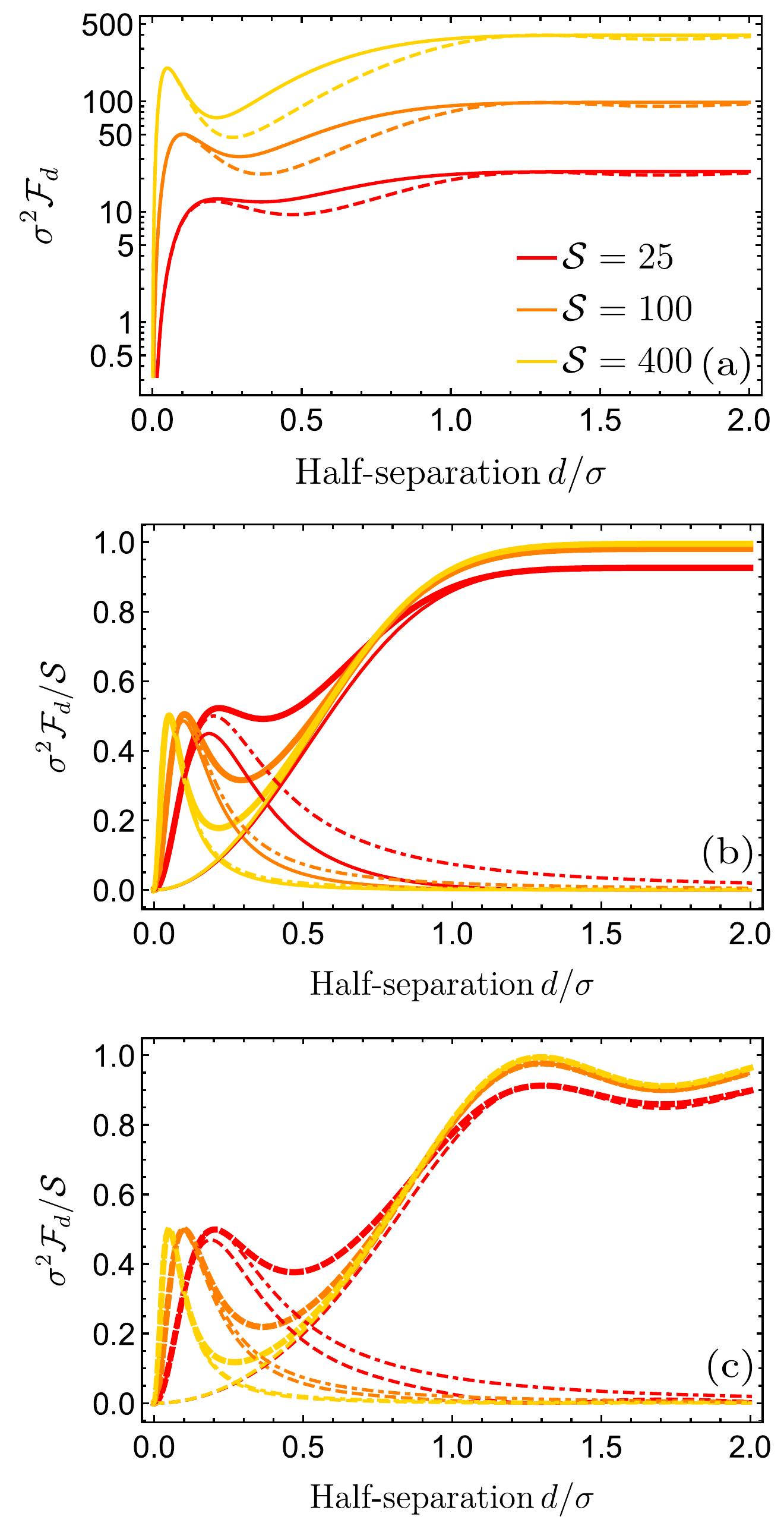}
\caption{(Color online) (a) The Fisher information $\sigma^2{\cal F}_d$ for estimating half-separation $d$ from the quadrature vectors $\mathbf{q}$ as a function of $d/\sigma$ for the SNR $\snr=25$ (dark grey, red online), $100$ (grey, orange online), $400$ (light grey, yellow online) and the soft (solid lines) and hard (dashed lines) aperture model. (b, c) The rescaled Fisher information $\sigma^2 {\cal F}_{d} /\snr$ (thick lines) for estimating  the half-separation shown as a sum of the Rayleigh $\mathcal{F}_d^{(\textrm{R})}$ and the sub-Rayleigh $\mathcal{F}_d^{(\textrm{SR})}$ parts (thin lines). Solid and dashed lines represent respectively the soft (b) and the hard (c) aperture model. The thin dashed-dotted line in panels (b) and (c) is analytical approximation of the sub-Rayleigh part given in Eq.~(\ref{Eq:FdSapprox}). Three values of the signal-to-noise ratio $\snr=25,100,400$ have been used with the same color coding as in the panel (a).}
\label{Fig:SeparationPrecision}
\end{figure}

Fig.~\ref{Fig:SeparationPrecision}(a) depicts the FI ${\cal F}_d$ for estimating the half-separation $d$ from the array homodyne measurement of field quadratures   for the SNR $\snr=25, 100$, and $400$,
calculated using Eqs.~(\ref{Eq:F(1)jj'})--(\ref{Eq:F(3)jj'}) in the limit $\Delta x \rightarrow 0$.
In the sub-Rayleigh region, when $d \ll \sigma$, a non-trivial feature appears in the form of a peak whose maximum shifts towards lower $d$ with increasing SNR. Interestingly, for high SNR the peak shape does not depend noticeably on the model of the transfer function.

Numerical calculations indicate that the sub-Rayleigh feature stems from the dependence on $d$ of the eigenvalues $V_\pm$ of the covariance matrix $\mathbf{C}$, which contributes to the Fisher information term ${\mathcal F}^{(1)}_{dd} $ given by Eq.~(\ref{Eq:F(1)jj'}). Because of its physical meaning, we will denote is as $\mathcal{F}_d^{(\textrm{SR})} \equiv {\mathcal F}^{(1)}_{dd}$.  In the strongly sub-Rayleigh regime one can derive a simple analytical approximation for this term as follows. For small separations the overlap $\ovp$ defined in Eq.~(\ref{Eq:OverlapDef}) can be approximated up to the quadratic order in $d$ using the Taylor series expansion $u(x\pm d) \approx u(x) \pm d u'(x) + {\textstyle\frac{1}{2}}
d^2 u''(x)$ as:
\begin{align}
\ovp & = \int_{-\infty}^{\infty}
 u(x+d) u(x-d) \Der x \nonumber \\
& \approx 1 - d^2 \int_{-\infty}^{\infty}   \{ [u'(x)]^2 - u(x)u''(x) \} \Der x
\nonumber \\
& = 1 - 2d^2 \int_{-\infty}^{\infty}   [u'(x)]^2 \Der x = 1 - \frac{2d^2}{\sigma^2}.
\label{Eq:Overlap}
\end{align}
In the third line integration by parts has been carried out assuming that the first derivative $u'(x)$ vanishes in the limit $x\rightarrow \pm \infty$ and the definition of $\sigma$ from Eq.~(\ref{Eq:sigmadef}) has been used. For clarity, the calculation in Eq.~(\ref{Eq:Overlap}) has been presented taking $\ctr=0$. The above approximation inserted into Eq.~(\ref{Eq:Vpm}) yields
\begin{equation}
\label{Eq:V-V+approx}
V_- \approx \snr d^2 /\sigma^2+ 1, \qquad V_+ \approx \snr (1-d^2 /\sigma^2)+1.
\end{equation}
As shown in Appendix~\ref{App:B}, for high SNR the dominant contribution to $\mathcal{F}_d^{(\textrm{SR})}$ in the sub-Rayleigh region comes from the dependence of $V_-$ on $d$. A simple algebra based on the approximate expression given in Eq.~(\ref{Eq:V-V+approx}) yields:
\begin{equation}
\label{Eq:FdSapprox}
\mathcal{F}_d^{(\textrm{SR})} \approx
\frac{1}{2}\left( \frac{1}{V_-} \frac{\partial V_-}{\partial d}\right)^2
\approx
\frac{\snr}{\sigma^2} \, f \! \left( \frac{\sqrt{\snr}d}{\sigma}\right),
\end{equation}
where we have introduced
\begin{equation}
\label{Eq:f(t)}
f(t) = \frac{2 t^2}{(1+t^2)^2}.
\end{equation}
Note that the first expression in Eq.~(\ref{Eq:FdSapprox}) can be viewed as Eq.~(\ref{Eq:FIMNormal}) specialized to the univariate case of the variance $V_-$ and a single estimated parameter $d$.

In Fig.~\ref{Fig:SeparationPrecision}(b,c)
we compare the second approximate expression in Eq.~(\ref{Eq:FdSapprox}) with the rescaled FI $\sigma^2{\cal F}_d/\snr$ respectively for the soft and the hard aperture models. It is seen that the approximation given in Eq.~(\ref{Eq:FdSapprox}) reproduces rather accurately the shape of the sub-Rayleigh feature for high SNR. An elementary analysis of the function $f(t)$ defined in Eq.~(\ref{Eq:f(t)})
yields the maximum of the sub-Rayleigh peak at $d=\sigma/\sqrt{\snr}$ and the endpoints of the half-maximum interval located at $(\sqrt{2}\pm 1)\sigma/\sqrt{\snr}$.
The second non-vanishing contribution to the FI, also shown in Fig.~\ref{Fig:SeparationPrecision}(b,c), comes from the change of the eigenvectors ${\mathbf e}_\pm$ with the half-separation $\theta$ in the orthogonal subspace $\mathbf{P}_\perp$. This contribution dominates in the Rayleigh region, when $d \gtrsim \sigma$, and will be denoted
${\cal F}_d^{(\mathrm{R})} \equiv {\cal F}_{dd}^{(2)}$.
As shown in Appendix~\ref{App:B} the Rayleigh term $\mathcal{F}_d^{(\textrm{R})}$
for $d \ll \sigma$ exhibits behavior $\mathcal{F}_d^{(\textrm{R})} \sim \snr d^2/[(1+\snr^{-1})\sigma^4]$ with the proportionality factor of the order of one and for $d \gg \sigma$ saturates at
\begin{equation}
\label{Eq:FdRlargedMain}
\mathcal{F}_d^{(\textrm{R})} = \frac{\snr}{\sigma^2}\frac{1}{1+2\snr^{-1}}, \qquad \mbox{[large $d$]}
\end{equation}
independently of the model of the transfer function selected in Eq.~(\ref{Eq:trans_funs}).

\subsection{Centroid estimation}
\label{Sec:CentroidPrecision}

\begin{figure}[t!]
\centering
\includegraphics[width=0.9\columnwidth]{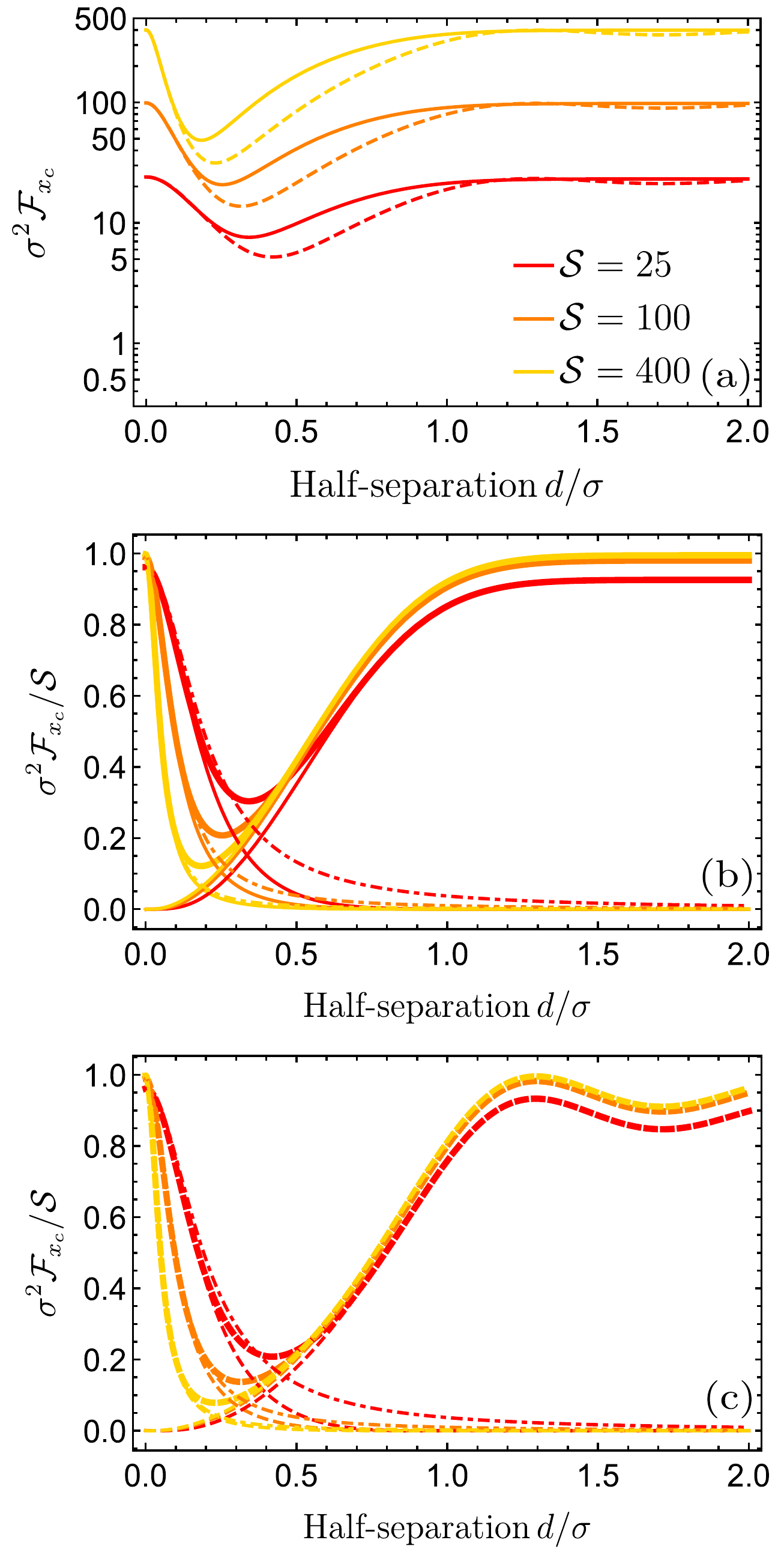}
\caption{(Color online) (a) The Fisher information $\sigma^2{\cal F}_\ctr$ for estimating the centroid $\ctr$ from the quadrature vectors $\mathbf{q}$ as a function of the half-separation $d/\sigma$ for the SNR $\snr=25$ (dark grey, red online), $100$ (grey, orange online), $400$ (light grey, yellow online) and the soft (solid lines) and hard (dashed lines) aperture model. (b, c) The rescaled Fisher information $\sigma^2 {\cal F}_{\ctr} /\snr$ (thick lines) for estimating  the centroid shown as a sum of the Rayleigh $\mathcal{F}_\ctr^{(\textrm{R})}$ and the sub-Rayleigh $\mathcal{F}_\ctr^{(\textrm{SR})}$ parts (thin lines). Solid and dashed lines represent respectively the soft (b) and the hard (c) aperture model. The thin dashed-dotted line in panels (b) and (c) is analytical approximation of the sub-Rayleigh part given in Eq.~(\ref{Eq:Fctrapprox}). Three values of the signal-to-noise ratio $\snr=25,100,400$ have been used with the same color coding as in the panel (a).}
\label{Fig:CentroidPrecision}
\end{figure}

Fig.~\ref{Fig:CentroidPrecision}(a) depicts the  FI ${\cal F}_\ctr$ for estimating the centroid $\ctr$ of a binary source as a function of the half-separation $d$ calculated using Eqs.~(\ref{Eq:F(1)jj'})--(\ref{Eq:F(3)jj'}). As detailed in Appendix~\ref{App:B}, the dip seen for $d \lesssim \sigma$ stems from the fact that ${\cal F}_\ctr$ is given by a sum of two contributions
$\mathcal{F}_\ctr = \mathcal{F}_\ctr^{(\textrm{R})} + \mathcal{F}_\ctr^{(\textrm{SR})}$ that dominate respectively in the Rayleigh and the sub-Rayleigh regions. These contributions are generated by the change of the eigenvectors $\mathbf{e}_{\pm}$ with the parameter $\ctr$ correspondingly in the subspace $\mathbf{P}_\perp$, producing $\mathcal{F}_\ctr^{(\textrm{R})} \equiv \mathcal{F}^{(2)}_{\ctr\ctr}$, and in the subspace spanned by $\mathbf{e}_\pm$, producing $\mathcal{F}_\ctr^{(\textrm{SR})} \equiv \mathcal{F}^{(3)}_{\ctr\ctr}$. The variances $V_\pm$ of the principal components do not depend on $\ctr$, hence
$\mathcal{F}^{(1)}_{\ctr\ctr} \equiv 0$.

As calculated in Appendix~\ref{App:B}, in the high SNR regime the Rayleigh part has the leading-order expansion around $d \approx 0$ in the form
$\mathcal{F}_\ctr^{(\textrm{R})} \sim \snr^2 d^4/\sigma^6$  with a proportionality constant of the order of one, while for $d \gg \sigma$ it approaches a constant value
\begin{equation}
\mathcal{F}_\ctr^{(\textrm{R})} = \frac{\snr}{\sigma^2}\frac{1}{1+2\snr^{-1}}, \qquad \mbox{[large $d$]}
\end{equation}
the same as in the case of estimating separation given in Eq.~(\ref{Eq:FdRlargedMain}).
This is easily understandable, as for large separations the locations of individual point sources can be determined independently and the half-separation and the centroid are respectively half of the difference and half of the sum of these locations.
As illustrated in Fig.~\ref{Fig:CentroidPrecision}(b,c), the sub-Rayleigh part $\mathcal{F}_\ctr^{(\textrm{SR})}$ is well approximated in the region $d \ll \sigma$ by an expression derived in Appendix~\ref{App:B}
\begin{equation}
\label{Eq:Fctrapprox}
\mathcal{F}_\ctr^{(\textrm{SR})} \approx \frac{\snr}{\sigma^2}\frac{1}{1+\snr^{-1}}
\left( 1 + \frac{\snr d^2}{\sigma^2}\right)^{-1}.
\end{equation}
This expression monotonically decreases from its maximum value $\snr/[\sigma^2(1+\snr^{-1})]$ at $d=0$,
reaching half-maximum at $d=\sigma/\sqrt{\snr}$. Note that at this value of $d$ the approximate form of
${\cal F}_d^{(\textrm{SR})}$ derived in Eq.(\ref{Eq:FdSapprox}) reaches its peak.

The non-trivial dependence of $\mathcal{F}_\ctr$ on the half-separation $d$ that results from combining $\mathcal{F}_\ctr^{(\textrm{R})}$ and $\mathcal{F}_\ctr^{(\textrm{SR})}$
can be explained intuitively as follows.
For the infinitesimal pixel size assumed here, the spatial intensity distribution obtained from
quadrature variances measured at individual pixels
is overwhelmed by the detection noise and cannot be used standalone for reliable estimation of the centroid. Instead, the information about the centroid is primarily obtained from coherences between pixels that induce non-trivial covariances between individual quadratures according to Eq.~(\ref{Eq:Covariance}). These coherences turn out to be most informative either when $d\approx 0$, i.e.\ one is effectively dealing with a single point source, or when the two points constituting the binary source are well separated.

\section{Estimation procedure}
\label{Sec:EstProcedure}

The analysis of the attainable precision carried out in the preceding section provides guidance to develop a practical algorithm for estimating the source parameters from the measured quadrature vectors.
Physically, the principal component $q_- = \mathbf{e}_-^T \mathbf{q}$ whose variance carries most of the information about the source separation in the sub-Rayleigh regime corresponds to the quadrature of the spatial mode given  in the image plane by $e_-(x)$ defined in Eq.~(\ref{Eq:epm(x)}). For small separations, $d \ll \sigma$, this mode function can be approximated using a straightforward Taylor series expansion by
\begin{equation}
\label{Eq:Approxe-(x)}
e_-(x) =
\frac{[u(x-\ctr-d)- u(x-\ctr+d)]}{\sqrt{2(1-\ovp)}}
\approx
v(x-\ctr),
\end{equation}
where $v(x)$ is the normalized derivative of the transfer function
\begin{equation}
\label{Eq:vdef}
 v(x) = -\sigma
 u'(x)
\end{equation}
that does not depend explicitly on $d$.
Note that in the second step in Eq.~(\ref{Eq:Approxe-(x)}) we have used the approximate form of the overlap $\ovp$ derived in Eq.~(\ref{Eq:Overlap}).

The variance $V_-$ that serves as the basis for estimating $d$ in the sub-Rayleigh region effectively measures the optical power carried in the spatial mode $e_-(x) \approx v(x-\ctr)$. This relates the estimation recipe for the source separation emerging from the principal component analysis to the currently explored approaches to superresolution imaging based on spatial mode demultiplexing (SPADE) \cite{TsangPRX2016} and similar techniques, where the separation of a binary source is inferred from the fraction of the optical power directed to carefully defined spatial modes in the image plane \cite{DuttonPRA2019}. For small separations, $d\ll \sigma$, the relevant information is contained predominantly in the optical power measured for the mode $v(x-\ctr)$ \cite{YinkIJQI2020}.
While SPADE and similar techniques require a careful alignment of the detection apparatus hardware with respect to the field in the image plane in order to avoid a systematic error \cite{ChrostowskiIJQI2017,RehacekPRA2017,arXiv:2003.01166}, the advantage of array homodyning is the ability to reconstruct the quadrature statistics for any relevant spatial mode through digital postprocessing of the quadrature vectors obtained from the pixelated measurement \cite{DawesPRA2003}. In particular, prior knowledge of the source centroid is not required to align the array homodyne detector. This is in contrast to SPADE-type techniques, where the need to determine the centroid results in an overhead in terms of the required signal \cite{Grace2020}. However, it needs to be verified whether the array homodyne data are sufficient on their own to estimate the separation in the sub-Rayleigh region. As we have seen in Sec.~\ref{Sec:CentroidPrecision}, the precision of estimating the source centroid exhibits a non-trivial dependence on the source separation.

\begin{figure}
\includegraphics[width=0.9\columnwidth]{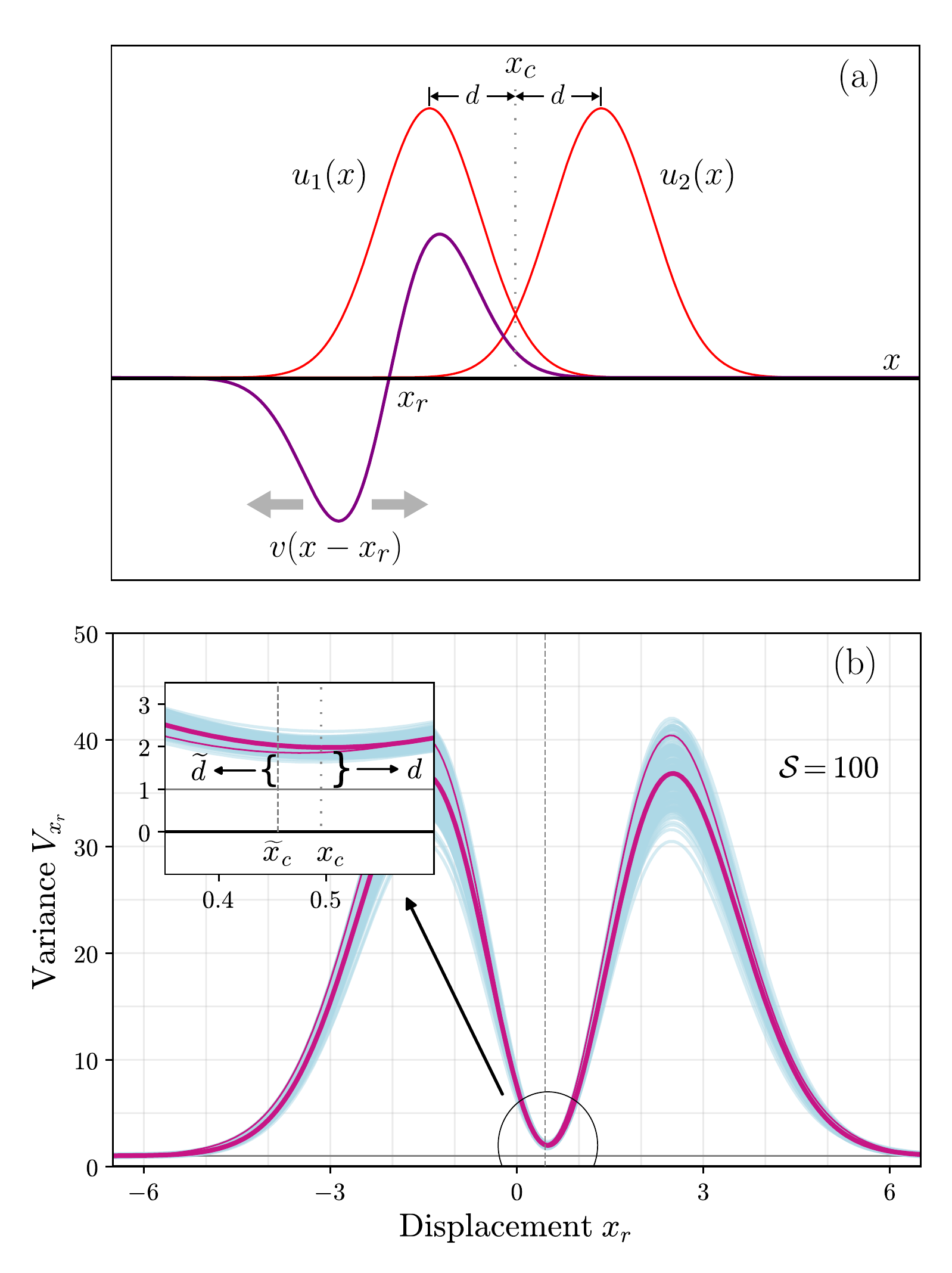}
\caption{(Color online) (a) Sweeping the mode $v(x-\xrf)$ along the image axis $x$ produces the variance $V_\xrf$ as a function of the displacement $\xrf$. (b) The graph of $V_\xrf$ (thick line) calculated for the binary source with a separation $d=0.2$, the soft aperture model with $\sigma=1$, and the SNR $\snr=100$. The graph is superposed on top of a collection of variance functions (thin light lines) calculated for individual realizations of a Monte Carlo simulated array homodyne detection experiment. The inset depicts schematically the estimation of the centroid $\tilde{x}_c$ and the separation $\tilde{d}$ from a single realization (thin dark line).}
\label{Fig:Vvxrf}
\end{figure}

The above question is answered positively by the following algorithm. Use the measured quadrature vectors $\mathbf{q}$ to determine the variance $V_{\xrf} = \mathrm{Var}[q_{\xrf}]$ of quadratures $q_{\xrf} = \sum_{i}  q_i v(x_{i}-\xrf)  \sqrt{\Delta x}$ defined for a one-parameter family of spatial modes obtained by displacing $v(x)$ specified in Eq.~(\ref{Eq:vdef}) by an arbitrary distance $\xrf$, as depicted in Fig.~\ref{Fig:Vvxrf}(a).
For a general discrete source, the variance $V_{\xrf}$ is given in the limit $\Delta x \rightarrow 0$ by an integral expression
\begin{equation}
V_{\xrf}
=  \snr \sum_{l} w_l \left| \int u_l^\ast(x) v(x-\xrf) \Der x\right|^2 + 1.
\label{Eq:Vvxrf}
\end{equation}
As shown in Fig.~\ref{Fig:Vvxrf}(b) for the soft aperture model, in the case of a binary source the graph of $V_{\xrf}$ as a function of $\xrf$ exhibits a two-lobe structure on top of the detection noise pedestal. The local minimum between the lobes can serve as an estimate for the centroid $\tilde{x}_c$. This facilitates the estimation of the source separation from the gap between $V_{\tilde{x}_c}$ and the detection noise level by inverting Eq.~(\ref{Eq:Vvxrf}) with $V_{\tilde{x}_c}$ used on the left hand side and $\ctr$ inserted in lieu of $\xrf$ on the right hand side. For $d\ll\sigma$, the estimation formula takes an approximate form $\tilde{d} \approx \sigma\sqrt{(V_{\tilde{x}_c}-1)/\snr}$ independently of the selected model of the transfer function.

\begin{figure}
\centering
\includegraphics[width=0.9\columnwidth]{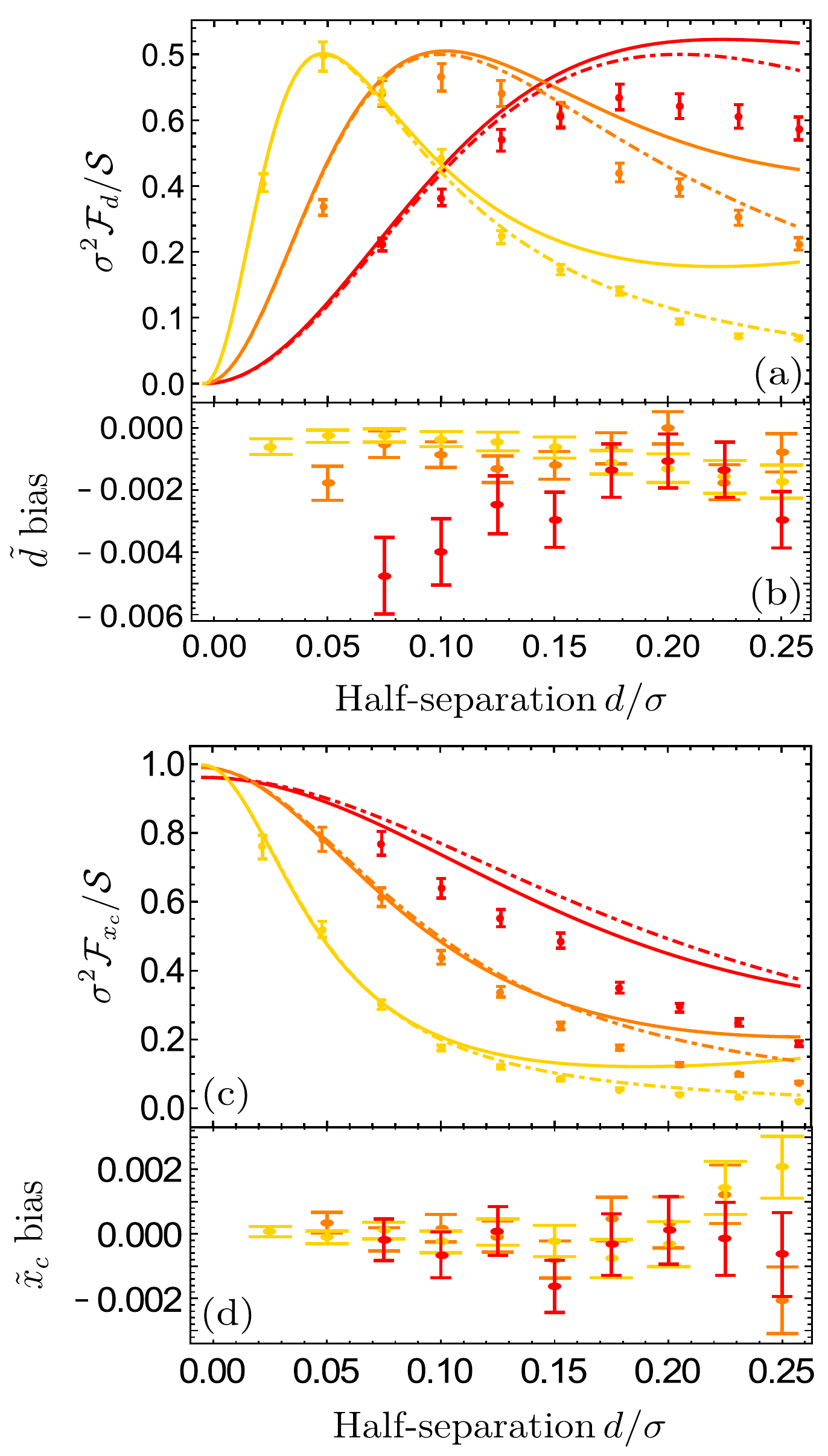}
\caption{(Color online) (a) Rescaled Fisher information $\sigma^2{\cal F}_d/\snr$ for estimating the half-separation $d$ in the sub-Rayleigh region for the soft-aperture model (solid lines) compared with the analytical approximation derived in Eq.~(\ref{Eq:FdSapprox}) (dash-dotted lines). The discrete points indicate precision of estimating $d$ from Monte Carlo data using the algorithm described in Sec.~\ref{Sec:EstProcedure}. The error bars correspond to the standard deviation of the precision normalized to the size $N$ of the sample. (b) The bias $\mathbbm{E}[\tilde{d}]-d$ of the estimator for the half-separation $d$. The error bars are taken as the standard deviation of the estimated value $\tilde{d}$. (c,d) Analogous graphs for estimation of the centroid $\ctr$ from the quadrature vector. Three values of the signal-to-noise ratio $\mathcal{S}=25, 100, 400$ have been used with the same color coding as in Fig. \ref{Fig:SeparationPrecision} and \ref{Fig:CentroidPrecision}.}
\label{Fig:MonteCarlo}
\end{figure}

The algorithm outlined above has been applied to Monte Carlo data generated by simulating for given source parameters 1000 realizations of the array homodyning experiment with the soft aperture model. In each realization, a sample of $N=500$ quadrature vectors has been drawn
for $1000$ pixels of width $0.008\sigma$ each. This sample was used to compute the variance $V_{\xrf}$ as a function of $\xrf$ and subsequently to determine from its local minimum the estimates for the centroid $\tilde{x}_c$ and the separation $\tilde{d}$. The precision of these estimates shown in Fig.~\ref{Fig:MonteCarlo}(a, c) is given by the squared inverse of the standard deviations for the histograms of $\tilde{d}$ and $\tilde{x}_c$ determined from individual realizations, after being rescaled by $N^{-1}$.
It is seen that for high SNR, the attainable precision follows the sub-Rayleigh part of the FI. However, it should be noted that the estimator $\tilde{d}$ for the half-separation exhibits a minor negative bias that can be observed in Fig.~\ref{Fig:MonteCarlo}(b). This is easily explained by the observation that estimating the half-separation $d$ from $V_{\ctr}$ would produce no bias, and replacing $V_{\ctr}$ by the local minimum value $V_{\tilde{x}_c}$ can only decrease the estimated value of $d$. In contrast, the estimator $\tilde{x}_c$ has no noticeable bias, as illustrated with Fig.~\ref{Fig:MonteCarlo}(d). Similar results have been obtained for Monte Carlo simulations using the hard aperture model.

\section{Conclusions}
\label{Sec:Conclusions}

\begin{figure}
\centering
\includegraphics[width=0.9\columnwidth]{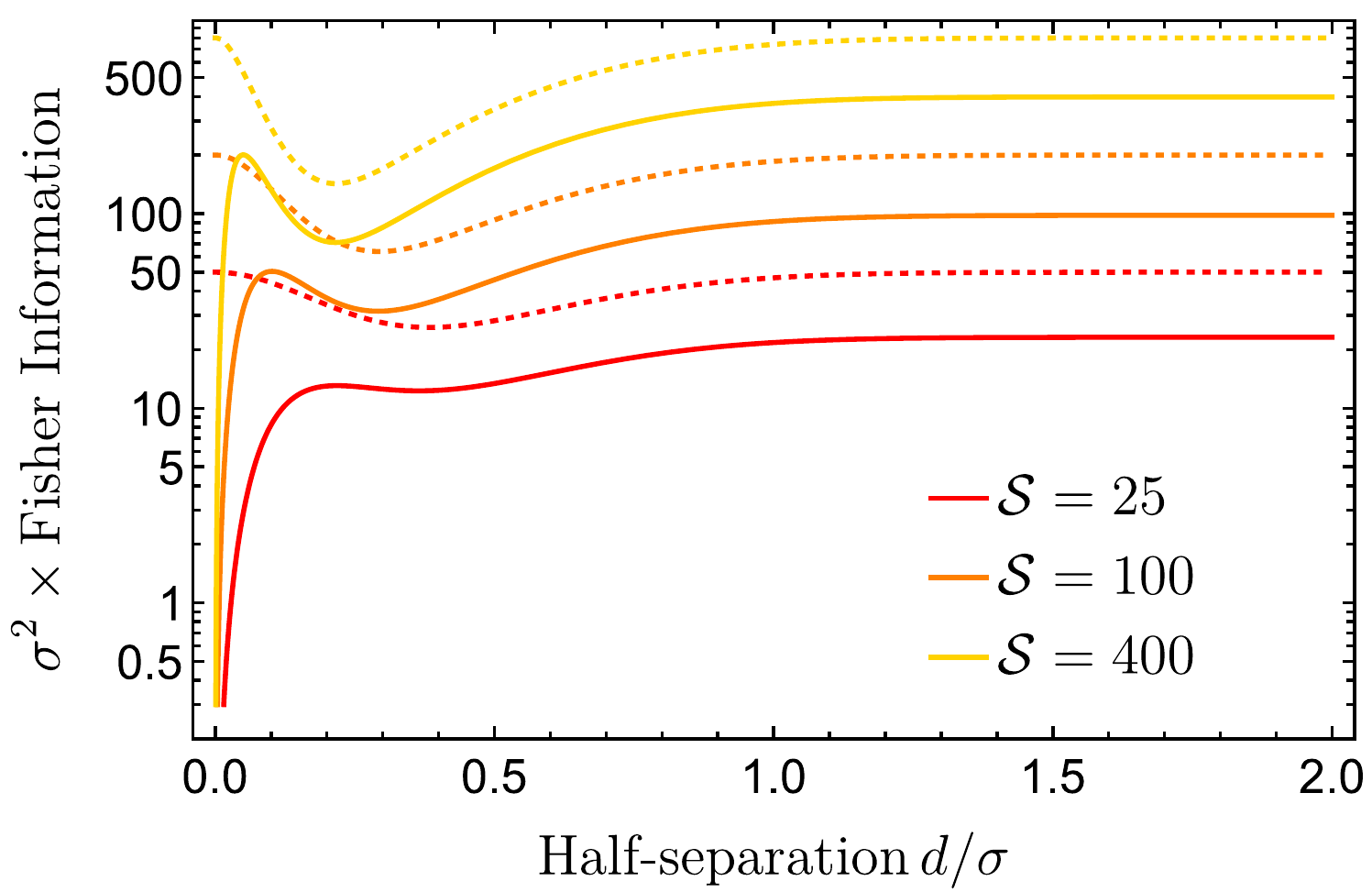}
\caption{(Color online) A comparison of the precision $\sigma^2 {\cal F}_d$ of estimating the half-separation $d$ from shot-noise limited array homodyne detection (solid lines) with the quantum mechanical bound assuming general measurements in the image plane (dotted lines). The signal-to-noise ratio $\snr$ is defined as two times the
the total average photon number reaching the image plane from the source. Soft aperture model has been used.}
\label{Fig:Quantum}
\end{figure}

Concluding, the analysis of estimating separation for a binary source demonstrates the potential of array homodyne detection to resolve spatial features of composite sources well below the Rayleigh diffraction limit, provided that sufficiently high SNR can be attained. In the case of shot-noise limited homodyne detection, the attainable precision can be directly compared with the ultimate quantum mechanical given by the quantum Fisher information \cite{NairTsang2016}, which takes into account the most general measurements in the image plane. In Fig.~\ref{Fig:Quantum} we compare the homodyne Fisher information ${\cal F}_d$ with its quantum mechanical counterpart for the soft aperture model, taking $\snr$ equal to two times the total average photon number reaching the image plane from the source. It is seen that for high SNR, or equivalently large photon number, the homodyne precision remains at approximately half of the quantum mechanical bound except in the direct vicinity of $d\approx 0$. The loss of precision of the homodyne measurement in this limit can be explained by the deleterious impact of the shot noise \cite{YinkIJQI2020}.

One should also note that array homodyning is only one of the available options for coherent detection in the image plane. Equivalent information about the source is obtained from the statistics of quadratures measured for orthonormal sets of spatial modes, e.g.\ Gauss-Hermite modes, that can be separated with the help of multiplane light conversion \cite{LabroilleOpEx2014}. Spatially multimode coherent detection is currently being developed as a technique to boost the capacity of optical communication links via spatial division multiplexing
\cite{RyfCLEO2019}. In a preliminary study following work presented here we have found that a joint measurement of both field quadratures enables one to handle complex transfer functions, albeit at a factor of two penalty for the SNR. Finally, an interesting extension of the presented work would be the reconstruction of properties of more intricate composite sources, also two-dimensional, using the spatial coherence information supplied by array homodyning.

\section{Acknowledgments}
We acknowledge insightful discussions with
K. Cha{\l}asi\'{n}ska-Macukow,
R. Demkowicz-Dobrza\'{n}ski, F. Ya. Khalili, and N. Treps.
This work is a part of the project ``Quantum Optical Technologies''
carried out within the International Research Agendas programme of the
Foundation for Polish
Science co-financed by the European Union under the European Regional
Development Fund. It was also supported by the US Department of Navy award no.\ N62909-19-1-2127 issued by the Office of Naval Research.

\appendix
\section{}
\label{App:A}

The inverse of the covariance matrix $\mathbf{C}$ given by Eq.~(\ref{Eq:Covform}) can be written explicitly as \begin{equation}
\mathbf{C}^{-1} = \sum_\mu V_\mu^{-1} \mathbf{e}_\mu \mathbf{e}_\mu^T + \mathbf{P}_\perp
\end{equation}
and its derivative with respect to the parameter $\theta_j$ reads
\begin{multline}
\frac{\partial {\mathbf{C}}}{\partial\theta_j} = \sum_\mu \left[ (\partial_{\theta_j} V_\mu) \mathbf{e}_\mu \mathbf{e}_\mu^T + \left(V_\mu-1\right) (\partial_{\theta_j} \mathbf{e}_\mu ) \mathbf{e}_\mu^T \right. \\
\left. + \left( V_\mu-1\right) \mathbf{e}_\mu  (\partial_{\theta_j} \mathbf{e}_\mu^T)\right].
\end{multline}
For the sake of brevity, we have introduced shorthand notation $\partial_{\theta_j} = \partial/ \partial\theta_j$. Using the above expressions,
the Fisher information matrix ${\boldsymbol{\mathcal F}}$ defined in Eq.~(\ref{Eq:FIMNormal}) can be conveniently written as a sum of three components ${\boldsymbol{\mathcal F}} = {\boldsymbol{\mathcal F}}^{(1)} + {\boldsymbol{\mathcal F}}^{(2)} + {\boldsymbol{\mathcal F}}^{(3)}$ with elements given by
\begin{widetext}
\begin{align}
{\cal F}^{(1)}_{jj'} & = \frac{1}{2}\sum_{\mu} \frac{1}{V_\mu^{2}} (\partial_{\theta_j} V_\mu) (\partial_{\theta_{j'}} V_\mu),
\label{Eq:F(1)jj'}\\
{\cal F}^{(2)}_{jj'} & = \sum_{\mu}\frac{(V_\mu-1)^2}{V_\mu}
(\partial_{\theta_j} \mathbf{e}_\mu)^T \mathbf{P}_\perp (\partial_{\theta_{j'}} \mathbf{e}_\mu)
= \sum_{\mu}\frac{(V_\mu-1)^2}{V_\mu}
\left( (\partial_{\theta_j} \mathbf{e}_\mu)^T (\partial_{\theta_{j'}} \mathbf{e}_\mu)
- \sum_{\mu'\neq\mu} (\partial_{\theta_j} \mathbf{e}_\mu)^T \mathbf{e}_{\mu'}
\mathbf{e}_{\mu'}^T (\partial_{\theta_{j'}} \mathbf{e}_\mu) \right),
\label{Eq:F(2)jj'} \\
{\cal F}^{(3)}_{jj'} & = \sum_{\mu} \sum_{\mu'\neq\mu}
\left(\frac{(V_\mu-1)(V_{\mu'}-1)}{V_\mu V_{\mu'}}[\mathbf{e}_\mu^T (\partial_{\theta_j} \mathbf{e}_{\mu'})][\mathbf{e}_{\mu'}^T (\partial_{\theta_{j'}} \mathbf{e}_\mu)]+\frac{(V_\mu-1)^2}{V_\mu V_{\mu'}}[\mathbf{e}_\mu^T (\partial_{\theta_j} \mathbf{e}_{\mu'})]
[\mathbf{e}_\mu^T (\partial_{\theta_{j'}} \mathbf{e}_{\mu'})]\right). \label{Eq:F(3)jj'}
\end{align}
\end{widetext}
The sums over $\mu'$ are restricted to $\mu' \neq \mu$ owing to the fact that
$\mathbf{e}_{\mu}^T (\partial_{\theta_j} \mathbf{e}_\mu)=0$, which follows directly from the normalization of the eigenvectors $\mathbf{e}_\mu$. Furthermore, the orthogonality of the eigenvectors implies that for any $\mu, \mu'$ one has
$\mathbf{e}_\mu^T (\partial_{\theta_j}  \mathbf{e}_{\mu'}) + \mathbf{e}_{\mu'}^T (\partial_{\theta_j}  \mathbf{e}_{\mu})
= \partial_{\theta_j} (\mathbf{e}_\mu^T\mathbf{e}_{\mu'})=0 $.

\section{}
\label{App:B}
When the covariance matrix given in Eq.~(\ref{Eq:Covform}) has only two distinguished eigenvectors, the third contribution ${\cal F}^{(3)}_{jj}$ to the diagonal elements of the Fisher information matrix given by Eq.~(\ref{Eq:F(3)jj'}) reduces to
\begin{equation}
\label{Eq:F(3)jj+-}
{\cal F}^{(3)}_{jj} = \frac{(V_+-V_-)^2}{V_+ V_-}
\left( \int_{-\infty}^{\infty} e_-(x) \partial_j [e_+(x)] \Der x \right)^2,
\end{equation}
where $j=d,\ctr$.
Specifically, for $j=d$ a straightforward calculation yields
\begin{equation}
\label{Eq:inte-partialde+}
\int_{-\infty}^{\infty} e_-(x) \partial_d [e_+(x) ]\Der x =
\int_{-\infty}^{\infty} e_+(x) \partial_d [e_-(x) ]\Der x =
0.
\end{equation}
This immediately implies that ${\cal F}_{dd}^{(3)}=0$.
Among the two remaining contributions to the Fisher information for separation estimation, ${\cal F}_{dd}^{(1)}$  is the sub-Rayleigh part ${\cal F}^{(\textrm{SR})}_{d} \equiv {\cal F}_{dd}^{(1)}$ discussed in detail in Sec.~\ref{Sec:SeparationPrecision}. For high signal-to-noise ratio, $\snr \gg 1$, the term in ${\cal F}_{dd}^{(1)}$ dependent on $V_+$ scales in the leading order of $d$ as $\frac{1}{2} (\partial_d V_+)^2/V_+^2 \approx 2\snr d^2/\sigma^4$. Hence it can be neglected in the region of small $d$ where the sub-Rayleigh feature occurs. The second contribution to ${\cal F}_{dd}$, referred to in Sec.~\ref{Sec:SeparationPrecision} as the Rayleigh part ${\cal F}_d^{(\mathrm{R})} \equiv {\cal F}_{dd}^{(2)}$, can be simplified using Eq.~(\ref{Eq:inte-partialde+}) to the form
\begin{multline}
\label{Eq:Fd(R)}
{\cal F}_d^{(\mathrm{R})} \equiv {\cal F}_{dd}^{(2)} = \frac{(V_+-1)^2}{V_+} \int_{-\infty}^{\infty}
   [\partial_d e_+(x)]^2 \Der x
   \\ + \frac{(V_--1)^2}{V_-} \int_{-\infty}^{\infty}
   [\partial_d e_-(x)]^2 \Der x.
\end{multline}
For small $d$ the leading contribution comes from the first term with $(V_+ -1)^2/V_+ \approx {\snr}/{(1+\snr^{-1})}$ and the integral $\int_{-\infty}^{\infty}
   [\partial_d e_+(x)]^2 \Der x$ producing an expression quadratic in $d$,
\begin{multline}
{\cal F}_d^{(\mathrm{R})} \equiv  {\cal F}_{dd}^{(2)}  \approx \frac{\snr d^2}{1+\snr^{-1}}
\left(  \int_{-\infty}^{\infty}   [u''(x)]^2 \Der x - \frac{1}{\sigma^4} \right). \\
\mbox{[small $d$]}
\end{multline}
The factor within the large round brackets is equal to $2/{\sigma^4}$ for the soft aperture model and $4/(5\sigma^4)$ for the hard aperture model. For large separations the overlap $\chi$ vanishes and the eigenvalues $V_\pm \approx \snr/2+1$ become equal to each other. Hence one can equivalently take as the eigenmodes in Eq.~(\ref{Eq:Fd(R)}) the displaced transfer functions $u(x-\ctr \pm d)$ instead of $e_\pm(x)$. The result is:
\begin{multline}
\label{Eq:FdRlarged}
{\cal F}_d^{(\mathrm{R})} \equiv  {\cal F}_{dd}^{(2)}  \approx  2 \frac{(\snr/2)^2}{1+\snr/2} \int_{-\infty}^{\infty} [u'(x)]^2 \Der x
\\ = \frac{\snr}{1+2 \snr^{-1}}\frac{1}{\sigma^2}. \qquad \mbox{[large $d$]}
\end{multline}
The physical interpretation is that in the regime of large separations, the half-separation is obtained through estimation of the locations $x_1$ and $x_2$ of the two peaks produces by individual points in the binary source and calculating $d=(x_1-x_2)/2$. The same argument holds in the limit of large separations for the contribution
${\cal F}^{(2)}_{\ctr\ctr}$ that plays the role of the Rayleigh part ${\cal F}_\ctr^{(\mathrm{R})}$ in the case of estimating the centroid $\ctr=(x_1+x_2)/2$. Hence a calculation analogous to the one leading to Eq.~(\ref{Eq:FdRlarged}) yields also
\begin{equation}
{\cal F}_\ctr^{(\mathrm{R})} \equiv {\cal F}_{\ctr\ctr}^{(2)}  \approx \frac{\snr}{1+2 \snr^{-1}}\frac{1}{\sigma^2}. \qquad \mbox{[large $d$]}
\end{equation}
Furthermore, a direct expansion into a power series shows that for  high signal-to-noise ratio, $\snr \gg 1$, and small $d$ the Rayleigh part has the leading-order term of the form ${\cal F}_\ctr^{(\mathrm{R})} \equiv {\cal F}_{\ctr\ctr}^{(2)} \approx \snr^2 d^4/\sigma^6$. The remaining part, given by Eq.~(\ref{Eq:F(3)jj+-}) specialized to $j=\ctr$, plays a non-trivial form in the sub-Rayleigh region and has been denoted
as ${\cal F}_\ctr^{(\mathrm{SR})} \equiv {\cal F}^{(3)}_{\ctr\ctr}$. Its approximate form for high signal-to-noise ratio $\snr \gg 1$ can be obtained by neglecting terms of the order of $d^2/\sigma^2$ in comparison to one, but retaining terms of the order of $\snr d^2/\sigma^2$. Under these assumptions one has $\bigl(\int_{-\infty}^{\infty} e_-(x) \partial_\ctr [e_+(x)] \Der x
\bigr)^2\approx 1/\sigma^2$, $V_+ \approx \snr + 1$, and $V_- \approx \snr d^2 / \sigma^2 +1$. Consequently,
\begin{equation}
{\cal F}_\ctr^{(\mathrm{SR})} \equiv {\cal F}^{(3)}_{\ctr\ctr}
\approx \frac{(V_+ - V_-)^2}{V_+ V_-} \frac{1}{\sigma^2},
 \qquad \mbox{[small $d$]}
\end{equation}
which after simplification yields Eq.~(\ref{Eq:Fctrapprox}) in Sec.~\ref{Sec:CentroidPrecision}.


\end{document}